# Double Slit Experiment from Nano to Femto Scale

*Arvind Khuntia[1] and Raghunath Sahoo[2]*

The evolution of light theories began with Isaac Newton's corpuscular model, which explained reflection and refraction but could not account for diffraction and interference. In contrast, Christiaan Huygens proposed a wave theory, explaining light's behavior through an ether-based medium, supported by his principle that each point in a wavefront acts as a secondary source. This idea was experimentally supported in the early nineteenth century when Thomas Young's double-slit experiment revealed an interference pattern, affirming light's wave nature. Later, James Clerk Maxwell unified electricity and magnetism, establishing light as an electromagnetic wave and extending the electromagnetic spectrum beyond visible light. In the twentieth century, Einstein's photoelectric effect introduced the concept of wave-particle duality, demonstrating that light behaves as discrete photons. Soon after, Louis de Broglie extended the idea of wave-particle duality to matter, a prediction confirmed in 1927 when Clinton Davisson and Lester Germer observed electron diffraction from a crystal and, independently, G.P. Thomson demonstrated electron diffraction through thin films, both proving that electrons also exhibit wave-like properties. This concept was dramatically visualized by Claus Jönsson's 1961 electron double-slit experiment. Recently, the ALICE collaboration observed quantum interference patterns at the femtometer scale in ultra-relativistic nuclear collisions, pushing quantum interference studies to new frontiers.

***Key Words:*** *UPC, photoproduction, interference, quark-gluon plasma, high-energy physics, double-slit experiment*

---

[1] Arvind Khuntia is at INFN, Bologna, Italy (email: Arvind.Khuntia@cern.ch)
[2] Raghunath Sahoo is at the Indian Institute of Technology Indore, India, Corresponding author email: Raghunath.Sahoo@cern.ch

In the seventeenth century, Isaac Newton proposed the corpuscular theory of light, suggesting that light consists of tiny particles or "corpuscles" emitted by luminous objects as they travel in a straight line. This theory explained reflection and refraction; however, it failed to account for phenomena like diffraction and interference. At the same time, Christiaan Huygens proposed the wave theory of light, which suggests that light travels as waves through an invisible medium called the ether. Huygens formulated his principle, which stated that every point on a wavefront acts as a source of secondary waves, helping to explain light's behaviour during reflection and refraction. In the early 19th century, Thomas Young performed the double-slit experiment [1], showing that light passing through two slits creates an interference pattern, an indication of its wave nature. Augustin-Jean Fresnel further developed this wave theory by mathematically describing light as a transverse wave, explaining phenomena like diffraction and polarization, which Newton's particle model failed to demonstrate. In the 1860s, James Clerk Maxwell revolutionised the understanding of light by unifying electricity and magnetism, demonstrating that light is an electromagnetic wave travelling through space at a finite speed. Maxwell's equations confirmed the wave nature of light and expanded the electromagnetic spectrum beyond visible light to include radio waves, microwaves, and X-rays.

The early twentieth century saw a major shift with Albert Einstein's explanation of the photoelectric effect, which introduced the concept of light quanta (photons). Einstein showed that light can also behave as particles, depending on the interaction, leading to the concept of wave-particle duality. This duality, confirmed by quantum mechanics experiments, demonstrated that even particles like electrons exhibit wave-like properties. A decisive breakthrough came in 1927, when Clinton Davisson and Lester Germer, in the United States, observed diffraction of electrons by a nickel crystal [2], providing direct evidence of the wave nature of electrons. Around the same time, George Paget Thomson in the United Kingdom independently demonstrated electron diffraction using thin metal foils [3]. These landmark experiments confirmed Louis de Broglie's hypothesis [4] that matter has wave properties and earned Davisson and Thomson the Nobel Prize in Physics in 1937. The double-slit experiment with electrons was first conducted by Claus Jönsson, a German physicist, in 1961. His work provided direct experimental evidence of electron wave-particle duality, confirming that individual electrons, when passed through two slits, produce an interference pattern, even when fired one at a time. This experiment built on Young's earlier double-slit experiment with light and expanded its implications by demonstrating that even fundamental particles like electrons exhibit wave-like behaviour under certain conditions. Experiments of this type generally produce interference patterns on the nanometer scale.

Imagine doing the famous double slit experiment, but instead of using two slits for particles to pass through, you use just a single atom. That is what scientists have now done with rubidium atoms and clever laser tricks [6]. They use two different colored lasers to excite the atom in two different ways. Each laser gives the electron inside the atom a different path to escape, but both paths lead to electrons coming out with exactly the same energy. This is like setting up two different doors for the electron to exit from, even though there is only one atom. When only one laser shines, there is just one door open, and the electrons do not show any pattern. When both lasers shine, the electron can leave by either door, and the waves of possibility from both paths overlap and interfere, creating a special pattern in the way electrons are detected.

The challenge was that these electrons have very little energy, so the scientists needed very sensitive equipment to see them clearly and to make sure nothing else disturbed them. What they found was that the electrons behave as waves until the very last moment when they are measured as particles, just as quantum theory predicts. So, with just a single atom and no slits, the same kind of interference that amazed scientists for centuries can be created in the tiniest possible way. This experiment not only confirms the strange nature of quantum particles but also opens up new ideas for exploring atoms and developing future quantum technologies.

In a recent study, the ALICE collaboration measured an interference pattern at the femtometer scale through ultra-peripheral collisions of Pb-nuclei at the Large Hadron Collider (LHC) [5].

***Young's double slit experiment in optics:***

Young suggested that light behaves as a wave, introducing the concept of a luminiferous ether permeating space to transmit these waves. He performed experiments to observe the interference patterns that emerge when light waves intersect, which highlighted light's wave-like behavior. Utilizing a double-slit setup, Young demonstrated that light passing through two nearby slits results in patterns of bright and dark fringes on a screen, a clear indication of wave interference. This observation challenged the prevailing particle theory of light, as particles would not produce such patterns. Young also demonstrated that the spacing of these fringes was influenced by the wavelength of the light, the separation between the slits, and the screen's distance from the slits. His experiments conclusively supported the wave theory, showing that light waves can constructively and destructively interfere, which results in bright and dark fringes, respectively. This phenomenon is clearly illustrated in Figure 1. When a monochromatic light source illuminates the double slit, the light waves passing through each slit overlap and interact. At certain points on the screen, the waves from both slits arrive in phase and reinforce each other, creating the bright fringes. At other points, the waves arrive out of phase and cancel each other out, producing the dark fringes. As a result, a series of evenly spaced vertical bright and dark bands, known as an interference pattern, appears on the screen. The precise spacing of these bands depends on the wavelength of the light, the distance between the slits, and the distance from the slits to the screen. The formation of this interference pattern, as seen in Figure 1, provides direct visual evidence that light exhibits wave-like behavior and can undergo constructive and destructive interference.

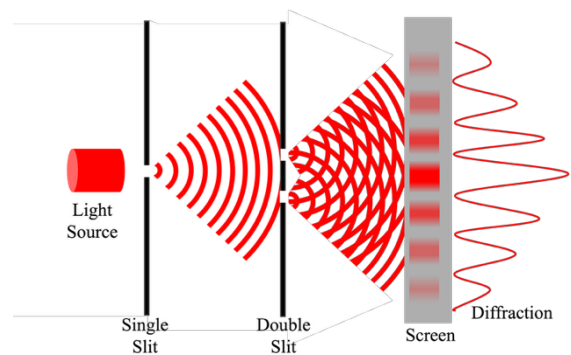

**Figure 1:** Schematic of Young's double-slit experiment

### Double slit experiment using electrons: wave nature of electrons:

Imagine you have a wall with two slits in it, and you start throwing balls at the wall. Most balls will hit the wall and bounce back, but some will go through the slits. If there's another wall behind the first one, the balls that pass through the slits will hit it. If you mark all the spots where a ball has landed on the second wall, you would expect to see two bands of marks, each lining up with one of the slits. This outcome makes sense because each ball is a solid object and can only go through one slit at a time, so it creates two clear strips of impact directly behind the slits. However, if you repeat this experiment with electrons instead of balls, the outcome is remarkably different. When a beam of electrons passes through the two slits, they do not simply create two bands on the screen as would be expected for ordinary particles. Instead, as illustrated in Figure 2, the electrons produce a series of alternating bright and dark bands known as an interference pattern, which is similar to the pattern observed with light waves [7].

This observation shows that electrons, although they are particles, also behave like waves. The probability of finding an electron at any point on the screen is determined by the way its associated wave passes through both slits at once and interferes with itself. At positions where the waves from the two slits reinforce each other, the probability is high and bright bands appear. At positions where the waves cancel each other out, the probability is low and dark bands appear. Even if electrons are sent one by one, over time this pattern emerges, which confirms the quantum nature of matter. The electron double slit experiment provides clear evidence for wave-particle duality. Electrons act as particles when detected, but their journey from source to screen is governed by wave-like interference, resulting in patterns that cannot be explained by classical physics alone.

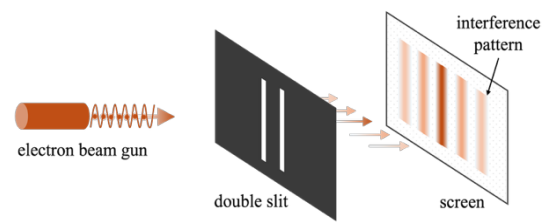

**Figure 2:** Schematic of the electron double-slit experiment. An electron beam gun emits electrons toward a barrier with two closely spaced slits. After passing through the double slit, the electrons form an interference pattern of alternating bright and dark bands on the detection screen, demonstrating the wave-like nature of electrons.

*UPC and double slit in femto-scale:*

The LHC is the largest and most powerful particle accelerator in the world. It began operation on September 10, 2008, and it features a 27-kilometer

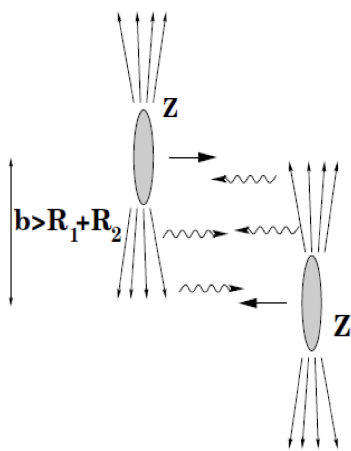

**Figure 3:** Schematic diagram of an ultraperipheral collision of two heavy-ions, where the impact parameter, b, is larger than the sum of the two radii [8].

ring of superconducting magnets. The LHC carried out collisions of heavy-ions to explore the behavior of strongly interacting matter under extreme temperatures and energy densities. These conditions give rise to a unique phase of matter known as quark-gluon plasma. In such collisions. The heavy nuclei are accompanied by a strong, Lorentz-contracted electromagnetic field, which acts as a flux of quasi-real photons. This provides the opportunity to study the photoproduction interactions at the LHC. The strength of the electromagnetic field surrounding a nucleus grows rapidly with the square of its charge ($Z^2$), meaning that heavy nuclei like lead produce much stronger electromagnetic fields than lighter elements. This makes photon-induced interactions highly likely in heavy-ion collisions. At the LHC, such interactions are studied through Ultra Peripheral Collisions (UPCs), where the impact parameter (the transverse distance between the two colliding nuclei) is larger than the sum of their radii, as shown in Figure 3. In such processes, the strong interactions are suppressed due to their short-range nature (O ~ few fm[a]), hence allowing one to separate photon-induced processes from hadronic interactions. Photonuclear production of vector mesons[b] received great interest among scientists, where the incoming photon fluctuates into a quark–antiquark pair that interacts strongly with the other incoming nucleus, resulting in the emission of a vector meson and the exchange of two gluons. Recently, ALICE has measured the interference pattern at the

---

[a] Fermi (fm): 1 fermi = $10^{-15}$ meter
[b] Vector meson: A vector meson is a spin-1 particle made of a quark and an antiquark (q$\bar{\text{q}}$), such as the ρ or ϕ.

femtometre scale using ultra-peripheral collisions between lead nuclei at the LHC using the coherently photoproduced ρ⁰ mesons through their di-pion decay channels. In coherent photoproduction, where the emitted photon interacts with the target nucleus as a whole, the latter remains intact and creates ambiguity as to which nucleus emits the photon and which serves as the target in the interaction, as shown

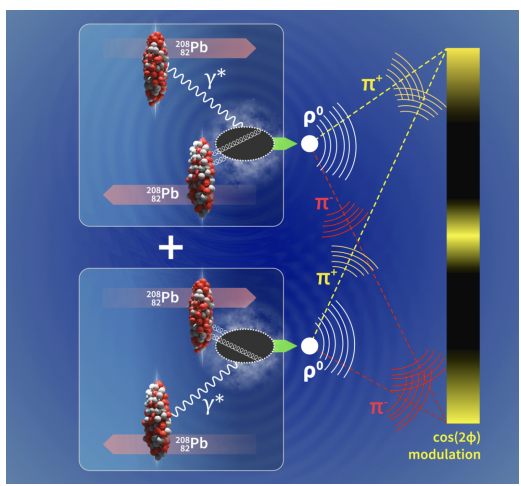

**Figure 4:** Schematic diagram of ρ0 vector meson production in the ultra-peripheral collision of Pb ions at the LHC and their decay through the di-pion channel. Courtesy: CERN CDS.

in Figure 4. The physics origin of this interference pattern at the femtometer scale comes from the fundamental principles of quantum mechanics. In an ultra-peripheral collision, either nucleus can emit the photon or act as the target, and there is no way to determine which scenario actually occurred. Because both possibilities are allowed by quantum mechanics, they both happen simultaneously as a quantum superposition. The probability waves for each scenario overlap and interfere, similar to how light waves overlap in Young's double slit experiment. This interference leads to a distinct pattern in the directions and momenta of the mesons detected by the experiment. The separation between the nuclei, known as the impact parameter, sets the effective distance between the "slits" in this analogy and is on the order of a few tens of femtometers. When the nuclei pass closer to each other, the interference becomes stronger and the pattern is more visible. In essence, the fact that we cannot tell which nucleus played which role means that quantum interference naturally appears, showing that the same quantum rules apply to massive nuclei as to light and electrons,

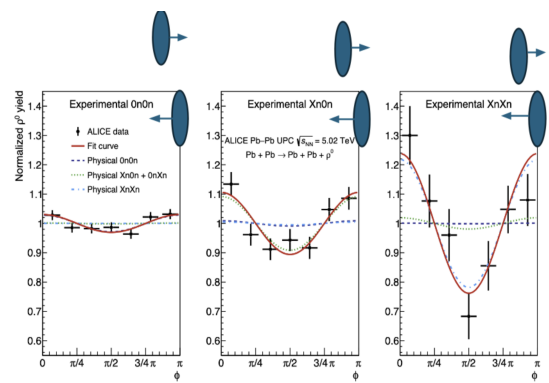

**Figure 5:** Normalized yield of photoproduced ρ⁰ mesons versus azimuthal angle φ, defined by the sum and difference of the pion transverse momenta, shown for three neutron emission classes (no emission, one-sided, and both-sided) [5].

just at a scale that is millions of times smaller.

In these studies, the photoproduced ρ⁰ vector meson inherits the linear polarization of the photon. This polarization, together with quantum interference, results in a correlation between the momentum and polarization of the $\rho^0$. As a result, the angular distribution of its decay products (two pions) exhibits a characteristic $\cos(2\phi)$ asymmetry, where $\phi$ is defined as the angle between the sum and the difference of the transverse momenta of the pions. This anisotropy, shown in Figure 5, becomes more pronounced as the impact parameter between the nuclei decreases, providing a clear signature of quantum interference at the femtometer scale.

***Summary and future directions:***

From Newton's particles to Huygens' and Young's wave theory, the nature of light and matter has been revealed through a series of foundational experiments, culminating in the modern understanding of quantum interference. The double slit experiment, first with light and later with electrons, established the principle that waves and even particles can interfere and create patterns that defy classical intuition.

Recent measurements at the LHC by the ALICE collaboration [5] have extended this quantum interference picture to the femtometer scale. By studying ultra-peripheral collisions of heavy nuclei, where the ambiguity of which nucleus emits the photon and which acts as the target leads to quantum interference, scientists have observed interference patterns in the production and decay of vector mesons. This effect, which appears as a modulation in the angular distribution of decay products, confirms that quantum mechanics governs matter at all scales, from photons and electrons to massive nuclei. The continuing study of these phenomena promises to deepen our understanding of the quantum structure of matter and the forces at play in the heart of atomic nuclei. With the much larger amount of data coming from Run 3 at the LHC, scientists will be able to study these quantum interference patterns more precisely than ever before. This will help reveal new details about how the building blocks of matter behave inside atomic nuclei. Such discoveries will bring us even closer to understanding the most fundamental laws of nature.

***Acknowledgements:*** The authors would like to thank Suraj Prasad, IIT Indore, for helping in the making of the figures.